\documentstyle[12pt]{article}
\begin{document}
\titlepage

\vfill
\newpage
\begin{center}
{\Large\bf Can the Fundamental Theory of Everything be Renormalizable?}

\medskip
\medskip
    {\bf J.\ Gegelia and N.\ Kiknadze}

\medskip

{\it High Energy Physics Institute and Dept.\ of General Physics,
Tbilisi State University, Chavchavadze ave.\ 1, Tbilisi 380028, Georgia.}
\end{center}

\medskip
\begin{abstract}
Some considerations showing that renormalizable theories with consistent
perturbative series can not be nonperturbatively finite (in terms of bare
parameters) are provided. Accordingly any fundamental unified theory
has to be either nonrenormalizable or order by order finite.
\end{abstract}

Renormalization procedure
serves well for extracting physical information from renormalizable
theories. Wilson's renormalization group approach \cite{wilson}
has contributed
much to deeper understanding of renormalizability. However,
despite the considerable success of renormalizable quantum field
theories a lot of physicists feel uneasy by necessity to deal with
divergent expressions. Some even declare that such QFT-s are
completely inconsistent \cite{dirac}. It is understood that any
self-contained theory, and hence fundamental unified theory of
all interactions too, must be nonperturbatively finite in terms of
bare parameters. In this
light, efforts to find order by order finite theories (main hopes
are relied on supersymmetric theories \cite{fin})
seem quite natural. There is a general feeling that divergences in
e.g.\ QED appear because QED itself is a low energy limit of unified
theory, and that correct treatment of gravity would
allow us to find self-consistent unified theory (see e.g.\
\cite{nakanishi}).
However, one may believe that divergences in
renormalizable theories are just artefact of perturbative approach
and the exact (nonperturbative) renormalization constants are finite
in terms of bare parameters. Below we are going to
demonstrate that this kind of viewpoint is not realistic. Although
this result is not a surprise, it was firmly established only for
superrenormalizable theories \cite{glimm}.

Let us consider some renormalizable theory and assume for a moment
that divergences are due only to perturbation theory, i.e. exact
expressions of physical quantities in terms of bare parameters are
finite. Let us employ dimensional regularization \cite{'t}. Relation
between bare ($g_0$) and renormalized ($g_\Lambda$) coupling constants
has the form:
\begin{equation}
g_0=\sum_{i=1}^{\infty} a_i g_\Lambda^i\ . \label{1}
\end{equation}                               
Here $\Lambda$ is the normalization point and divergences inhabit
coefficients $a_i$. Our assumption of the finiteness of exact
solutions implies that (\ref{1}) is a formal expansion of some finite
(in $\epsilon\to 0$ limit) relation:
\begin{equation}
g_0=f(g_\Lambda,\epsilon)\stackrel{\epsilon\to 0}{\longrightarrow}
g_\Lambda Z_{exact}(g_\Lambda)\ ,     \label{2}
\end{equation}                                        
with $Z_{exact}$ being finite. Of course existence of the zero
$\epsilon$ limit in (\ref{2}) does not imply that this limit necessarily
exists for the coefficients of its expansion in powers of $g_\Lambda$.

Alternately one could use the MS scheme \cite{hooft}. Then
expression (\ref{1}) takes the form:
\begin{equation}
g_0=\mu^\epsilon \left(g_{MS}+\sum_{i=1}
b_i(g_{MS})\epsilon^{-i}\right)\ .   \label{3}
\end{equation}
The important point here is that coefficients $b_i$ are independent
of 't Hooft's unit mass $\mu$ as well as of other dimensional
parameters. Relation between $g_\Lambda$ and $g_{MS}$ is given by
series with some finite coefficients:
\begin{equation}
g_{MS}=g_\Lambda+\sum_{i=3}^{\infty} c_i g^i_\Lambda\ .  \label{4}
\end{equation}                             
Suppose (\ref{3}) is a formal expansion of some finite
(in the $\epsilon\to 0$ limit) function (evidently, it is impossible
in superrenormalizable theory, where there are only finite number of
diverging terms in (\ref{3})):
\begin{equation}
g_0=\mu^\epsilon \phi^*(g_{MS},\epsilon)\ . \label{5}
\end{equation}                          
Taking limit $\epsilon\to 0$ in (\ref{5}) we get
\begin{equation}
g_0=\phi(g_{MS})                        \label{6}
\end{equation}    
and hence $g_{MS}$ does not depend on $\mu$ (here $\phi$, like
$Z_{exact}$ in (\ref{2}), is defined up to a function with
zero asymptotic expansion).
It implies vanishing of $\beta_{MS}$ function. On the other
hand, in perturbation theory for renormalizable models
$\beta_{MS}\neq 0$ and certain relations between $b_i$ in (\ref{3})
guarantee order by order finiteness of it \cite{hooft}. As far as
there exist no other asymptotic expansion of zero then with coefficients
identically equal to zero, we see that the asymptotic character of
the series in minimal schemes is not compatible with the finiteness
of the exact solutions of the theory. One may claim
that the minimal schemes are inconsistent (i.e.\ (\ref{4}) is not
asymptotic even if the renormalized series in $g_\Lambda$ are).
However, the same kind of analysis holds in any particular scheme.
Consider, for example,
regularization $\Lambda$, which was absent initially.
Differentiation of the expression
$g_0=Z\left(\Lambda,g(\Lambda)\right)g(\Lambda)$ with respect to
$\Lambda$ can be employed in derivation of series for $\beta$-function
(example of such calculation can be found e.g.\ in \cite{itz}).
Order by order finiteness of resulting series for $\beta$-function
in the limit when regularization is removed is related to the
renormalizability of the theory. If the theory were nonperturbatively
finite, the perturbative series for $Z$ would represent expansion
of some $Z_{exact}$, independent from $\Lambda$ in the removed
regularization limit and hence leading to zero exact
$\beta$-function.

  Our results for $MS$ scheme evidently agrees with exact solutions for
bare parameters by 't Hooft \cite{hooft}.

Note that our argument holds only if differentiation with respect to
mass scale commutes with ``summing" of perturbation series. But if
it were not the case, then perturbation series would have nothing to
do with the exact expressions. Although there is no theoretical
proof of the asymptotic character of perturbation series for physically
interesting renormalizable theories (QED, for example), it is
anticipated due to the success of them in describing experimental
data.

So we have demonstrated that if renormalized perturbative series in
renormalizable theory have any status (i.e.\ are asymptotic), then
nonperturbative relations between bare and renormalized quantities are
necessarily divergent.
Hence, it is clear that any candidate for fundamental unified theory
must be either order by order finite or perturbatively nonrenormalizable
by standard approach. If taking (supposedly) low-energy limit in such
theory leaves
us with the standard model, it means that relations between bare and
renormalized quantities diverge in that limit, while relations
between renormalized quantities remain finite and they can be
extracted by renormalization procedure.

\end{document}